\documentstyle[12pt]{article}
\title{\huge  A Monte Carlo study of random surface field effect on layering transitions}
\author{\bf
 H. Ez-Zahraouy$^{*}$, L. Bahmad, and A. Benyoussef
\\
 Laboratoire de Magn\'{e}tisme et de la Physique
 des Hautes Energies
\\
Universit\'{e} Mohammed V, Facult\'{e} des Sciences, Avenue Ibn Batouta,\\
Rabat  B.P. 1014, Morocco
}
\date{ }
\begin{document}
\maketitle

\begin{abstract}
\mbox{~~~}
The effect of a random surface field, within the bimodal distribution, on the
layering transitions in a spin-1/2 Ising thin film  is
investigated,  using Monte Carlo simulations. It is found that the
layering transitions depend strongly on the 
concentration $p$ of the disorder of the surface magnetic field, for a
fixed temperature, surface and external magnetic fields. Indeed,
the critical concentration $p_c(k)$ at which the magnetisation of each
layer $k$ changes the sign discontinuously, decreases for increasing
the applied surface magnetic field, for fixed values of the
temperature $T$ and the external magnetic field $H$.
Moreover, the behaviour of the layer magnetisations as well as the
distribution of positive and negative spins in each layer, are also
established for specific values of $H_s$, $H$, $p$ and the
temperature $T$. \\

\end{abstract}
----------------------------------- 
\mbox{~}(*) Corresponding author: ezahamid@fsr.ac.ma \\
\section{Introduction}
\mbox{  }
Several authors have studied the layering transitions of magnetic Ising systems,
such as  Pandit {\it et al.} [1], Pandit and Wortis
[2], Nightingale {\it et al.} [3] and Ebner {\it et al.} 
[4-7]. 
Many experimental studies have motivated these
theoretical works. Indeed, several methods have been 
used to study the layering transitions in Ising magnetic films. 
Benyoussef and Ez-Zahraouy have studied these layering transitions 
using a real space 
renormalization group [8], and transfer matrix methods [9].
Using the mean field theory, Hong [10], have found that depending on
the values of the exchange integrals near the surface 
region, the film critical temperature may be lower, higher than, or equal to that of the bulk. 
Using the perturbative theory, Harris [11] have showed the existence
of layering
transitions at $T=0$ in the presence of a transverse magnetic
field. The effect of finite size on such transitions has been studied,
in a thin film confined between parallel plates or 
walls, by Nakanishi and Fisher [12] using mean field theory and by
Bruno ${\it et al.}$ [13] taking into account the 
capillary condensation effect. 
By applying Monte Carlo simulations on thin Ising films with competing
walls, Binder {\it et al.} [14], found that 
occurring phase transitions belong to the universality class of the
two-dimensional Ising model and found that the 
transition is shifted to a temperature just below the wetting
transition of a semi-infinite fluid [15,16]. Hanke 
${\it et al.}$ [17] showed that symmetry breaking fields give rise to
nontrivial and long-ranged order parameter profiles 
for critical systems such as fluids, alloys, or magnets confined to
wedges. 
Much attention has been paid to the properties of layered
structures consisting of alternating magnetic materials. The most commonly
studied magnetic multilayers are those of ferromagnetic transition metal such
as $Co$ or $Ni$. Many experiments have shown that the magnetization enhancement
exists in multilayered films consisting of magnetic layers. It was found that
ferromagnetic coupling can exist between magnetic layers. From the
theoretical point of view, great interest has been paid to spin wave
excitations as well as critical phenomena [18-21]. The study of thin
films is partly motivated by the development of new 
growth and characterisation techniques, but perhaps more so by the
discovery of many exciting new properties, some quite unanticipated. These
include, more recently, the discovery of enormous values of magnetoresistance
in magnetic multilayers far exceeding those found in single layer films and
the discovery of oscillatory interlayer coupling in transition metal multilayers.
These experimental studies have motivated much theoretical work. However
these developments are applied to a large extent powered by "materials engineering"
and the ability to control and understand the growth of thin layers only a
few atoms thick. However, Many experimental studies have shown that
the magnetisation enhancement exists in multilayered 
films consisting of magnetic layers.
On the other hand, we have shown in one of our earlier works [22] the
existence of layering transitions under the effect of a variable
surface coupling. Moreover, for an Ising film with a wedge,
we found in Ref. [23] the intra-layering transitions under the
geometry effect consisting on the existence a wedge.
When the film is subject to a random transverse magnetic field [24],
we found the layer-by-layer transitions when increasing the
concentration $p$ above a critical value $p_c(k)$ for each layer $k$.
The aim of this work is to study the effect a random surface magnetic
field on layering transitions of an Ising thin film, using Monte Carlo simulations.
The paper is organised as follows. In section $2$, we describe the model
and the method used: Monte Carlo (MC) simulations. In section $3$ we present
results and discussions. 

\section{Model and Monte Carlo simulations}
We are studying a magnetic thin film formed with $N=4$ layers coupled ferromagnetically.
Each layer is a square of dimension $N_x \times N_y=64 \times 64$ spins.
$N_x$ and $N_y$ stand for the number of spins in the $x$ and $y$ directions,
respectively. \\
A preliminary study showed that, when performing Monte Carlo simulations under the Metropolis algorithm,
we note that the relevant calculated quantities did not
change appreciably for small film thickness: $N=3$, $N=4$, $N=5$
and $N=8$ layers; and when varying the number of spins of each layer from
$N_x=N_y=32$ to $128$. 
Taking into account the above considerations, in all the following we
will give  numerical
results for a film thickness $N=4$ layers, and $N_x=N_y=64$ spins
for each layer. \\
A surface magnetic field $H_s$ is acting only on spins
of the surface $k=1$. \\
The Hamiltonian governing the system is given by
\begin{equation}
{\cal H}=-J\sum_{<i,j>}S_{i}S_{j}-\sum_{i}H_{i}S_{i}
\end{equation}
where, $S_{l}(l=i,j)=-1,+1$ are the spin variables and the interaction
between different spins is assumed to be constant. The total magnetic
field $H_i$ applied on each site $i$ of the layer $k$, is
 distributed according to:
\begin{equation}
H_{i}=\left\{
	\begin{array}{crl}
	H+H_{si} &  \mbox{for all sites i $\epsilon$} & \mbox{surface k=1} \\
	H    &  \mbox{for all sites i $\epsilon$} & \mbox{layers k=2,3,4}.
	\end{array}
    \right.
\end{equation}
$H$ is an external magnetic field, and $H_{si}$ is the surface magnetic
field distributed according to a bimodal law: \\
\begin{equation}
{\cal P}(H_{si})=p\delta(H_s-H_{si})+(1-p)\delta(H_{si}),
\end{equation}
so that the surface magnetic field is:
\begin{equation}
H_{si}=\left\{
	\begin{array}{cc}
	H_s &  \mbox{with the probability p} \\
	0   &  \mbox{with the probability 1-p} \\
	\end{array}
    \right.
\end{equation}
where $0\le p \le 1$. For $p=1$, $H_s$ is acting on all sites of the
surface; whereas $p=0$ corresponds to a situation with no surface magnetic
field: $H_s=0$. \\
\section{Results and discussion}
 \mbox{~~~}
We present in Fig. 1 the ground state phase diagram in the plane $(H,H_s)$ for a film with
$N=4$ layers, with different line transitions and configurations. For
$H_s \ge H_{sc}=N/(N-1)$, the system transits from the
configuration $O^N$ to the configuration $1O^{N-1}$ at the line
$H=1-H_s$; and the configuration $1O^{N-1}$ transits to $1^{N}$ at a constant magnetic field
$H=-1/(N-1)$. 
Whereas the transition from the configuration $O^N$ to $1^{N}$ is
located at $H=-H_s/N$, provided that the surface magnetic field is
positive and kept less than its critical value: $H_s \le H_{sc}$. It is
worth to note that all these configurations are located at negative
values of the external magnetic field $H$.
In this work, the temperature fluctuations will not be taken into
account. Hence, the temperature will be fixed at the value 
$T=3.75$ in all the following.
In order to outline the effect of the surface magnetic field
concentration, we plot in Fig. 2 the phase diagram of the studied system in
the plane $(H,p)$, for $H_s=1.5$. We show the existence
of two critical probabilities: $p_L^{S}$ and $p_L^{F}$.
The former is the layering concentration corresponding to the
transition of the surface $k=1$, and located at about $ \approx
0.40$. The surface line transition is terminated by  the endpoint $C_1$ located at the
critical concentration $p_c(1) \approx 0.60$.
The latter is the film transition located at approximately  $ \approx
0.55$ as it is shown in the inset of Fig. $2$.
It is worth to note that the internal layers of the film transit at
different lines with different endpoints. Indeed, the second layer $k=2$
transits at line terminated by the endpoint $C_2$ located at $ \approx
0.94$, the third layer corresponds to the line with the endpoint  $C_3$ located at $ \approx
0.92$, whereas the last layer $k=4$ at the line with the endpoint
$C_4$ corresponding to the critical concentration $ \approx
0.88$. \\
These critical probabilities depend strongly on
the surface magnetic field values, at a fixed external
magnetic field as it illustrated by Fig. $3$. When increasing $H_s$, the
critical concentration of the surface $p_c(1)$ as well as for internal layers
$k=2,3,4$, $p_c(k)$ decreases  when increasing $H_s$ more and more. In all
cases the critical concentration of the surface $p_c(1)$ is always
found to be smaller than  $p_c(k), k=2,3,4$ of the other layers. This
is due to the fact that the surface transits, in all cases, before the
other layers. 
This is outlined in Fig. $4$, for a small concentration
$p=0.30$ of $H_s$ for a fixed surface magnetic field $H_s=1.5$.
To complete this study, we give in Figs. $5a,5b$ the distribution of
spins of the surface before and after the transition shown in
Fig. $4$, for $H_s=1.5$ and $p=0.30$. Fig. $5a$ shows the map of
negative spins for $H=-0.20$ before the transition, whereas 
Fig. $5b$ corresponds to positive spins for $H=-0.05$ after the
transition located at an external magnetic field $H \approx
-0.08$. This magnetic field value depends not only on the
concentration $p$ but also on the surface
magnetic field $H_s$; and decreases when increasing the concentration
$p$.
 Indeed, the random
distribution of $H_s$ is responsible on the formation of small
and randomly distributed islands of negative and positive spins present in the
surface and also in the deeper layers $k=2,3,4$. When $p
\rightarrow 1.0$ the negative spins,
or positive ones, are assembled on lager islands in a given layer and
the randomness of $H_s$ is lost. 

\section{Conclusion}
We have studied the effect of a random surface magnetic field, within the bimodal distribution, on the
 layering transitions of a spin-1/2 Ising thin film  formed with $N=4$
 layers, using Monte Carlo simulations. 
We showed that the layering transitions depend strongly on the 
concentration p of the surface magnetic field, for a
 fixed surface and external magnetic fields. Moreover,
 the critical concentration $p_c(k)$ of each layer $k$ decreases with
 increasing the applied surface magnetic field, for a fixed value
of the external magnetic field. 
In order to complete this study, we have investigated the
 magnetisation behaviour for each layer as a function of the external
 magnetic field for fixed values of $T$, $H_s$, and the concentration $p$.
We have also shown the existence of islands  of negative spins
 surrounded by aggregation of positive spins. \\
\\
\noindent{ \Large \bf References}
\begin{enumerate}

\item[{[1]}] R. Pandit, M. Schick and M. Wortis, Phys. Rev. B {\bf 26}, 8115 (1982).
\item[{[2]}] R. Pandit and M. Wortis, Phys. Rev. B {\bf 25}, 3226  (1982).
\item[{[3]}] M. P. Nightingale, W. F. Saam and M. Schick, Phys. Rev. B {\bf 30},3830 (1984).
\item[{[4]}] C. Ebner, C. Rottman and M. Wortis, Phys. Rev. B {\bf 28},4186  (1983).
\item[{[5]}] C. Ebner and W. F. Saam, Phys. Rev. Lett. {\bf 58},587 (1987).
\item[{[6]}] C. Ebner and W. F. Saam, Phys. Rev. B {\bf 35},1822 (1987).
\item[{[7]}] C. Ebner, W. F. Saam and A. K. Sen, Phys. Rev. B {\bf 32},1558 (1987).
\item[{[8]}]  A. Benyoussef and H. Ez-Zahraouy, Physica A, {\bf 206}, 196
(1994).
\item[{[9]}]  A. Benyoussef and H. Ez-Zahraouy, J. Phys. {\it I } France
{\bf 4}, 393 (1994).
\item[{[10]}] Q. Hong  Phys. Rev. B {\bf 41}, 9621  (1990);   ibid, Phys. Rev. B {\bf 46}, 3207  (1992).
\item[{[11]}] A. B. Harris, C. Micheletti and J. Yeomans, J. Stat. Phys. {\bf 84}, 323 (1996)
\item[{[12]}] H. Nakanishi and M. E. Fisher, J. Chem. Phys. {\bf 78},3279  (1983)
\item[{[13]}] P. S. Swain and A. O. Parry, Eur. Phys. J.  {\bf B 4}, 459 (1998);
E. Bruno, U. Marini, B. Marconi and R. Evans, Physica A {\bf 141A}, 187 (1987)
\item[{[14]}] K. Binder, D. P. Landau and A. M. Ferrenberg, Phys. Rev. Lett. {\bf 74}, 298 (1995)
\item[{[15]}] K. Binder, D. P. Landau and A. M. Ferrenberg, Phys. Rev. {\bf E 51}, 2823 (1995)
\item[{[16]}] M. Bengrine, A. Benyoussef, H. Ez-Zahraouy and F. Mhirech, Physica {\bf A 268}, 149 (1999)
\item[{[17]}] A. Hanke, M. Krech, F. Schlesener and S. Dietrich,
  Phys. Rev. E {\bf 60}, 5163 (1999)
\item[{[18]}] S. Dietrich and M. Schick, Phys. Rev B {\bf 31},4718 (1985)
\item[{[19]}] S. J. Kennedy and S. J. Walker, Phys. Rev. B {\bf 30},1498 (1984)
\item[{[20]}] P. Wagner and K. Binder, Surf. Sci. {\bf 175},421 (1986)
\item[{[21]}] K. Binder and D. P. Landau, Phys. Rev. B {\bf 37}, 1745 (1988)
\item[{[22]}] L. Bahmad, A. Benyoussef, and H. Ez-Zahraouy, Surf. Sci. {\bf 536}, 114 (2003)
\item[{[23]}] L. Bahmad, A. Benyoussef, and H. Ez-Zahraouy, Phys. Rev. E {\bf 66}, 056117 (2002)
\item[{[24]}] L. Bahmad, A. Benyoussef, and H. Ez-Zahraouy, Chin. J. Phys. {\bf 40}, 537 (2002)

\end{enumerate}
\noindent{ \Large \bf Figure Captions}\\

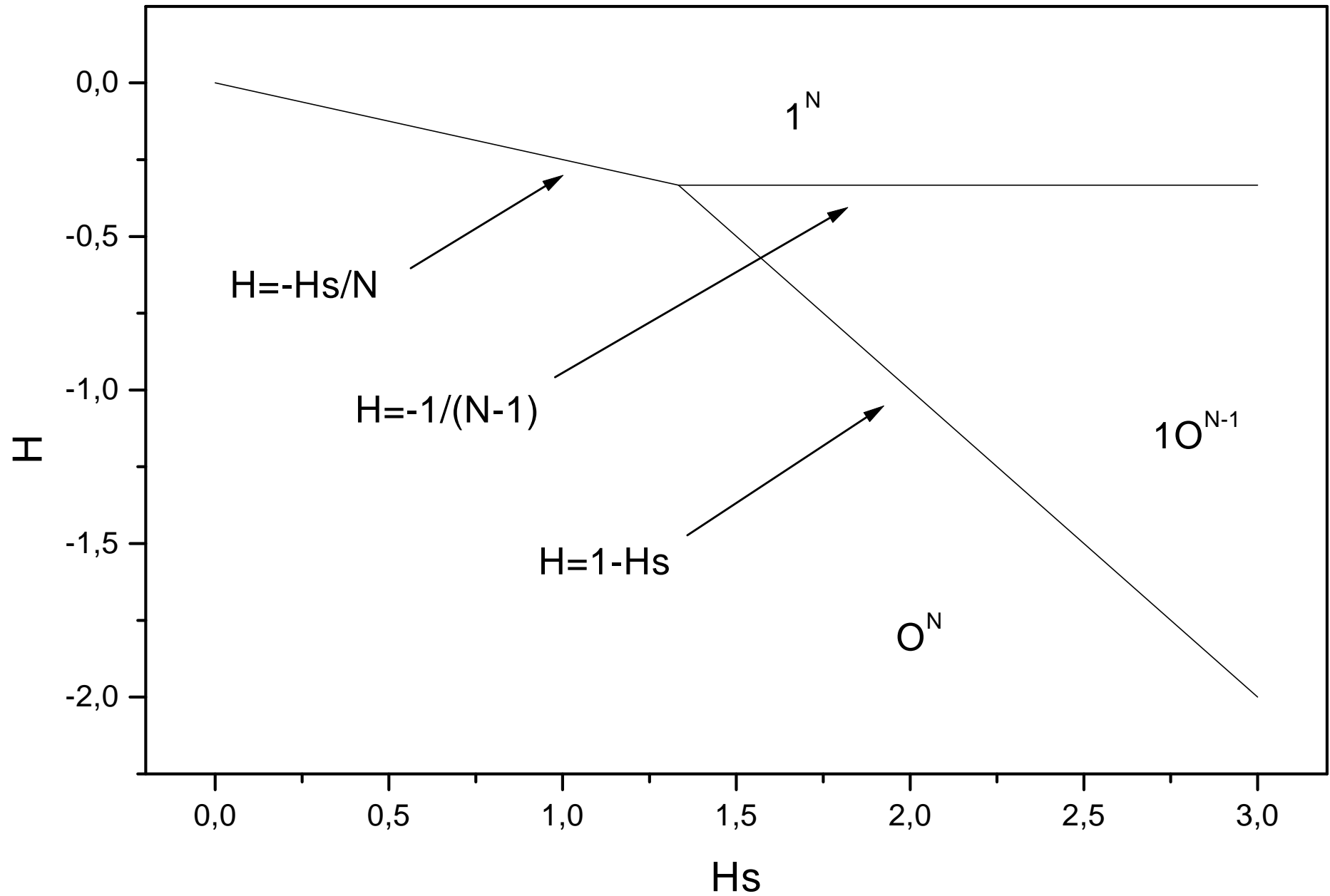
\noindent{\bf Figure 1.}: \\
The ground state phase diagram in the plane $(H,H_s)$ for a film with
$N=4$ layers, with different line transitions and configurations. \\

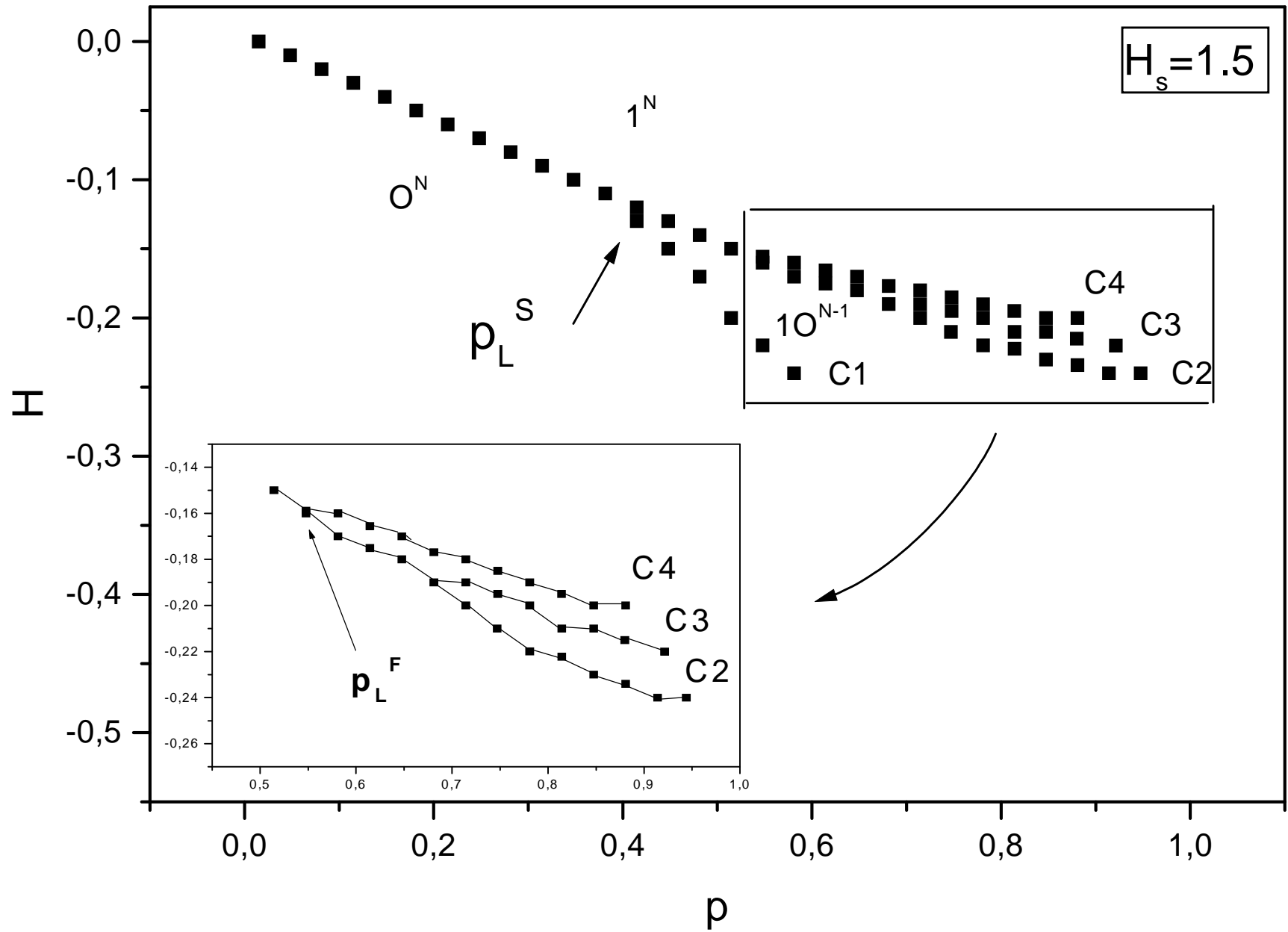
\noindent{\bf Figure 2}: \\
Phase diagram in the plane $(H,p)$ for a fixed temperature $T=3.75$ and a surface
magnetic field value $H_s=1.5$, showing the layering transition probability of
the surface $p_L^{S}$ as well as the endpoints $C_k, k=1,2,3,4$.
Inset, the layering transition probability of the film $p_L^{F}$ 
and the layering probability of the remaining
layers. \\

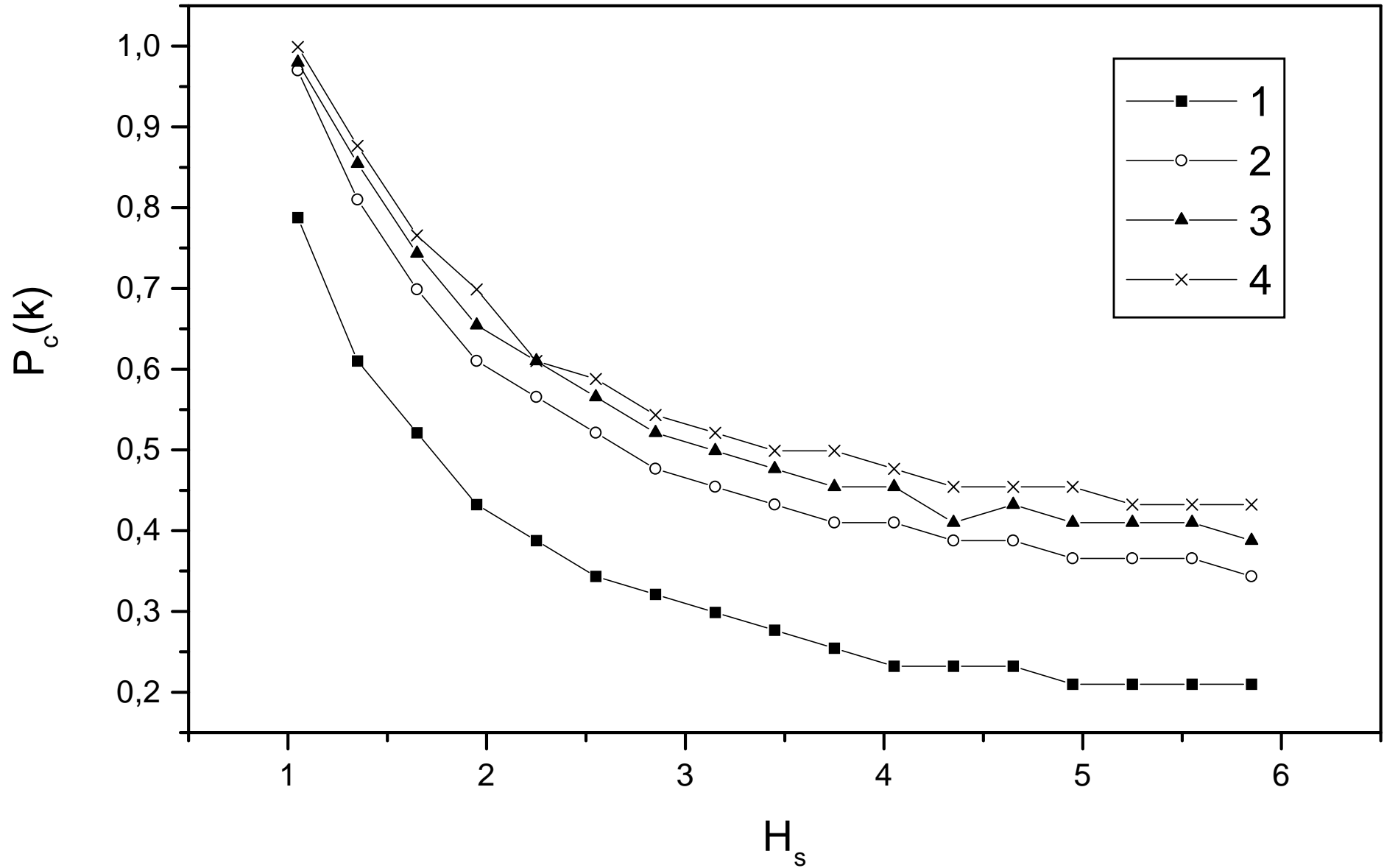
\noindent{\bf Figure 3}: \\
Critical probabilities $p_c(k)$ for each layer, as a function of the surface
magnetic field $H_s$ for an external magnetic field $H=-0.20$. \\

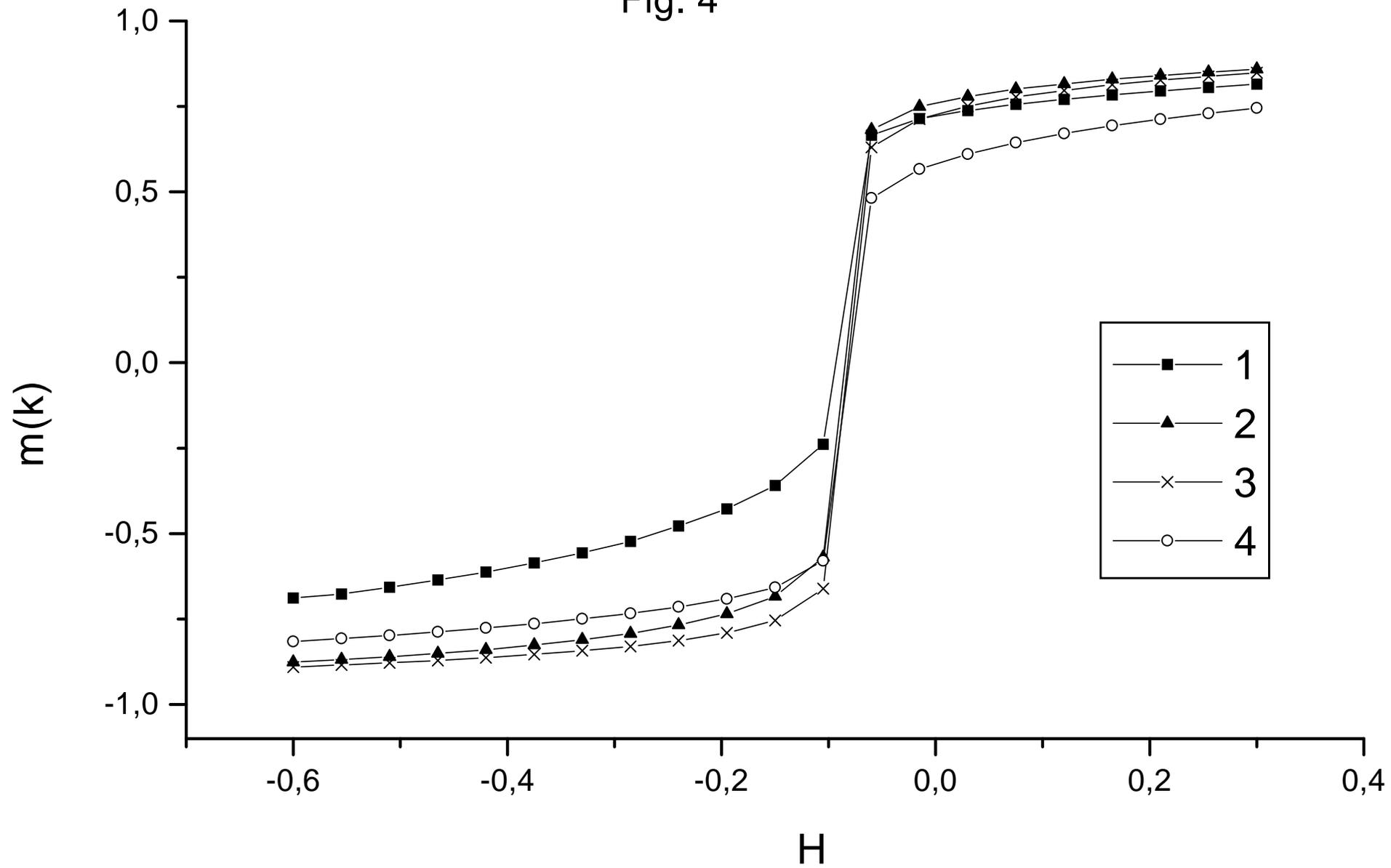
\noindent{\bf Figure 4}:\\
Magnetisation profiles as a function of $H$ for different layers $k=1,2,3,4$ of the film,  
for $H_s=1.5$ and a concentration value $p=0.30$. 
The number accompanying each curve is the layer order. \\

\noindent{\bf Figure 5}: \\
Distribution of spins of the surface $k=1$ before and after the transition
for a film with $N=4$ layers,  $T=3.75$, $H_s=1.5$ and $p=0.30$:
(a) negative spins for $H=-0.20$ before the transition, 
(b) positive spins for $H=-0.05$ after  the transition. 

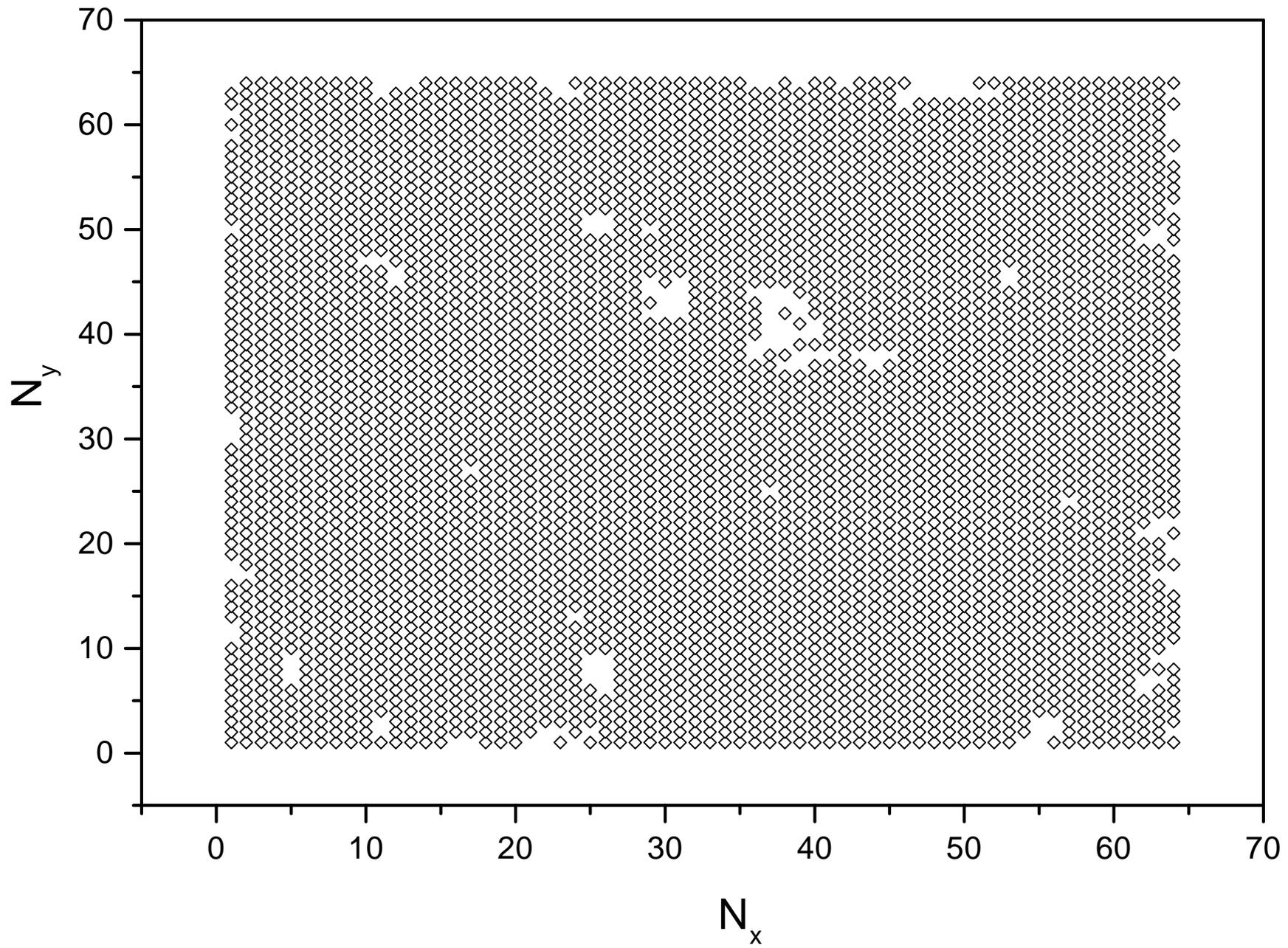
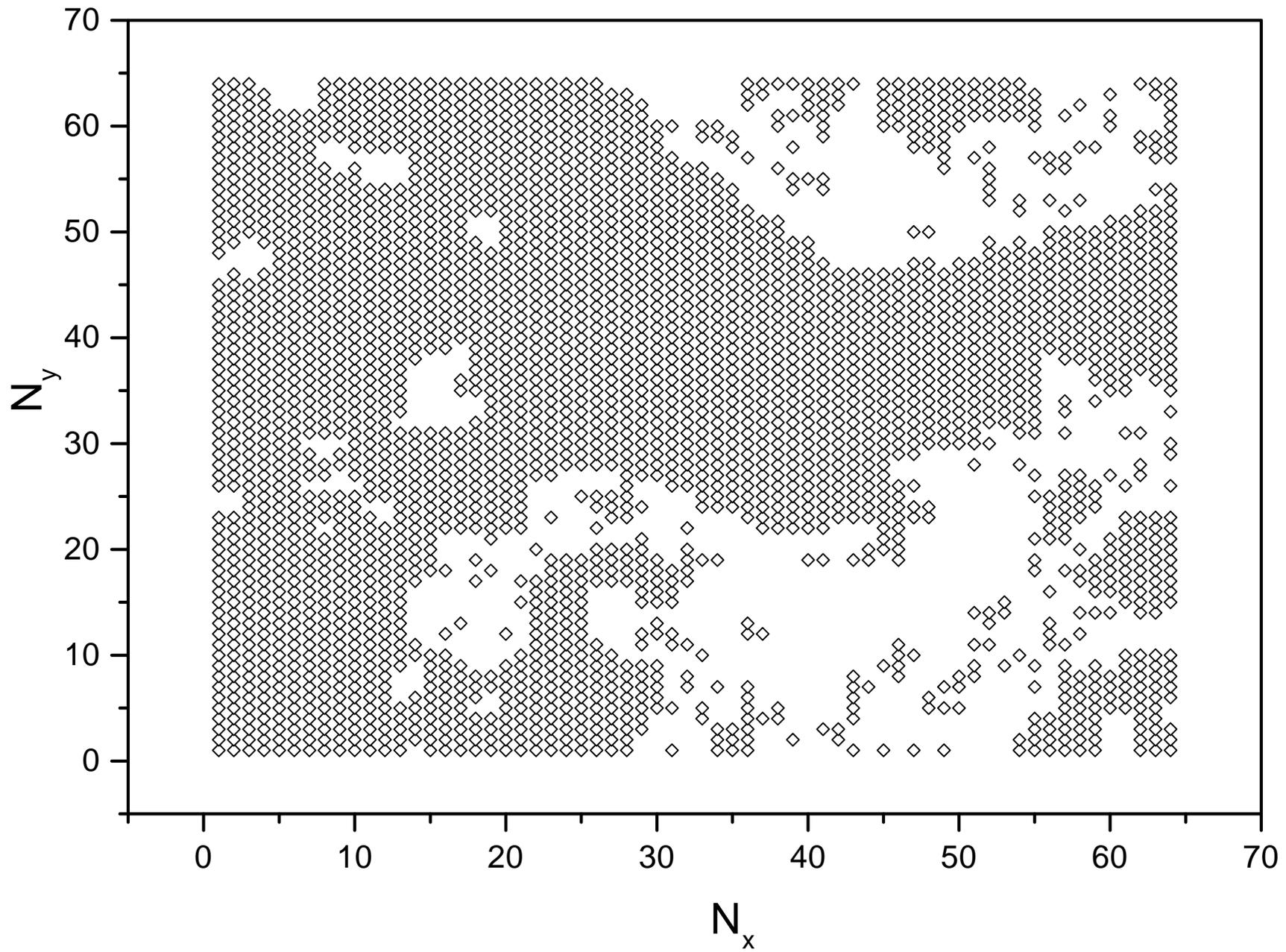
\end{document}